\documentclass[12pt]{article}
\usepackage{amsmath,amssymb,
}
\newtheorem{theorem}{Theorem}[section]
\newtheorem{proposition}[theorem]{Proposition}
\newtheorem{prop}[theorem]{Proposition}
\newtheorem{lemma}[theorem]{Lemma}

\newtheorem{defi}[theorem]{Definition}

\newcommand{\CH}{{\cal H}}

\newcommand{\CA}{{\cal A}}

\newcommand{\CB}{{\cal B}}

\newcommand{\CZ}{{\cal Z}}
\newcommand{\CS}{{\cal S}}

\newcommand{\CF}{{\cal F}}
\newcommand{\CR}{{\cal R}}
\newcommand{\CG}{{\cal G}}
\newcommand{\CM}{{ \cal M}}

\newcommand{\CL}{{\cal L}}

\makeatletter
\@addtoreset{equation}{section}

\makeatother
\newcommand{\Ee}{{\mathsf E}}

\def\al{\alpha}

\def\la{\lambda}

\newcommand{\jmp}[3]{Jour. Math. Phys. {\bf #1} (#2), #3}
\newcommand{\rref}[1]{(\ref{#1})} %puts parentheses around ref's

\def\dsl{\displaystyle}

\def\mat2#1#2#3#4{{\left(\begin{array}{cc}#1 & #2\\ #3 & #4
      \end{array}\right)}}

\def\mats2#1#2#3#4{{\left(\begin{array}{cc}#1 & #2\vspace{2truemm} \\ #3 & #4
\end{array}\right)}}

\def\ddd#1#2{\displaystyle{\frac{\partial #1}{\partial #2}}}

\newcommand{\Ha}[1]{H^{(#1)}}
\newcommand{\Ka}[1]{K^{(#1)}}

\def\Nij{Nijenhuis}

\def\endpf{\begin{flushright}$\square$\end{flushright}}

\def\var{manifold}

\def\bih{bi-Ham\-il\-tonian}
\def\varb{\bih\ \var}

\def\omnman{$\omega N$ manifold}

\def\dncoo{Darboux-Nijenhuis coordinates}

\def\St{St\"ackel}

\def\evalc{{\Big\vert_{\buildrel{\la=f}\over{\mu=g}}}}
\def\evzer{{\Big\vert_{{c_1=c_n=0}}}}

\def\bil{bi-Lagrangian}
\def\bilf{\bil\ foliation}
\newcommand{\SoV}{Separation of Variables}

\newcommand{\Lie}[1]{{\text{Lie}_{{#1}}}}
\def\bilf{\bil\ foliation}

\newcommand{\fino}[3]{{\>#1=#2,\ldots,#3}}
\begin{document}
\begin{flushright}
Ref. SISSA 29/2005/FM
\end{flushright}
%\vspace{0.8truecm}
\baselineskip=24pt
\begin{center}
{\Large\bf Gel'fand-Zakharevich Systems
and Algebraic Integrability: the Volterra Lattice Revisited}\end{center}
\vspace{0.5truecm}
%\bigskip
\begin{center}
\baselineskip=18pt
{\large Gregorio Falqui}\\
SISSA, via Beirut 2/4,
I--34014 Trieste, Italy,
falqui@fm.sissa.it\\
{\large Marco Pedroni}\\
Dipartimento di Ingegneria Gestionale e dell'Informazione,\\
Universit\`a di Bergamo,
Viale Marconi 5,
I-24044 Dalmine (BG), Italy,
marco.pedroni@unibg.it
\end{center}
\baselineskip=14pt
%{\bf Keywords:} {Hamilton-Jacobi Equations, Bihamiltonian Manifolds,
%Separation of Variables, Generalized Toda Lattices.}
\begin{abstract}\noindent
In this paper we will discuss some features of the \bih\ method
for solving the Hamilton-Jacobi (H-J) equations by Separation of
Variables, and make contact with the theory of Algebraic Complete
Integrability and, specifically, with the Veselov--Novikov notion
of {\em algebro-geometric (AG) Poisson brackets}. The \bih\ method
for separating the Hamilton-Jacobi equations is based on the
notion of pencil of Poisson brackets and on the
Gel'fand-Zakharevich (GZ) approach to integrable systems. We will
herewith show how, quite naturally, GZ systems may give rise to AG
Poisson brackets, together with specific recipes to solve the H-J
equations. We will then show how this setting works by framing
results by Veselov and Pensko\"\i\ about the algebraic
integrability of the Volterra lattice within the \bih\ setting for
Separation of Variables

\end{abstract}
%\tableofcontents
\section{Introduction}
The Hamilton-Jacobi equations, and the problem of their
separability, are one of the many fields in mathematics in which
the influence and heritage of Carl Gustav Jacob Jacobi
is still alive.
Such a problem, which can be considered one of the fundamental
problems of Theoretical Mechanics, is rooted in the foundational works of
Jacobi, St\"ackel, Levi-Civita and others. It has recently received a strong renewed
interest thanks to its applications to the theory of integrable PDEs of KdV
type (namely, the theory of finite-gap integration) and to the theory of
quantum integrable systems (see, e.g.,~\cite{DKN90,Sk95}).

The constructive definition of separability originally due to
Jacobi
%and which we will concentrate on in this paper,
is the following.

Let us consider an {\em integrable} Hamiltonian $H$ on a
$2n$-dimensional phase space, that is, let us suppose that, along
with $H=H_1$ we have further $n-1$ mutually commuting integrals of
the motion $H_2,\ldots H_n$, with $dH_1\wedge\ldots \wedge
dH_n\neq 0$.
\begin{defi}
An integrable system $(H_1,\ldots,H_n)$ is separable in the canonical coordinates
$(\mathbf{p},\mathbf{q})$ if there exist $n$ independent
relations
\begin{equation}
  \label{eq:1.sjask}
  \Phi_i(q_i,p_i;H_1,\ldots,H_n)=0,\quad i=1,\ldots n\>,
\end{equation}
connecting {\em single pairs $(q_i,p_i)$} of coordinates with the $n$
Hamiltonians $H_j$.
\end{defi}
The link of this definition with the theory of those integrable
systems that admit a Lax representation with spectral parameter
\[
\dot{L}(\la)= [L(\la),P(\la)],
\]
such as those associated with classical limits of quantum spin
systems and/or those coming from suitable reductions/restrictions
of KdV-like evolutionary PDEs, is self evident. Indeed, the Lax
representation of a system provides us with a natural candidate
for the separation relations: the characteristic polynomial of
$L(\la)$, also known as the {\em spectral curve} associated with
$L(\la)$. However, the possibility of successfully applying the
Lax method relies on three non-algorithmic steps to be solved:
\begin{itemize}
\item {\em To find} the Lax representation of a dynamical system;
\item To prove that the spectral invariants of $L(\la)$ are mutually in involution
(i.e., to prove that the Lax representation is compatible with a classical
$r$-matrix structure);
\item To give canonical coordinates as the coordinates of $n$ points
lying on the spectral curve, i.e., to actually implement what is
sometimes called {\em Sklyanin's magic recipe}.
\end{itemize}
%% DA QUI .......
The setting devised by Veselov and Novikov~\cite{VN85} to
characterize algebraic integrability requires that the phase space
$\CM$ of a Hamiltonian system fulfill some properties. They can
be, quite roughly, summarized as follows:
\begin{description}
\item{a)} $\CM$  has the fibered structure
\begin{equation}
\CM^{}{\buildrel{{S^k\Gamma}}\over{{\longrightarrow}}}\CB,
\end{equation}
where the base $\CB$ is a $n$-dimensional manifold whose points
determine an algebraic curve $\Gamma(b)$, and the fiber is the
$k$--th symmetric product of that curve. In more details, one
requires that $\Gamma(b)$ be given as an $m$--sheeted covering
$\Gamma(b){\buildrel{{m}}\over{{\longrightarrow}}} \mathbb{C}$ of
the complex $\la$-plane, and that points of $\CM$ can be
parameterized via the curve $\Gamma(b)$, and a set of $k$ points
on it, that is, the coordinates $\la_1$,\ldots,$\la_k$ of the
projection on the $\la$-plane of a set of points on it, as well as
discrete parameters $\epsilon_i$ that specify on which sheet of
the covering the points live.

\item{b)} An Abelian differential $Q(\Gamma)$ on $\Gamma$ (or
possibly on a covering of $\Gamma$), smoothly depending on the
points $b\in \CB$, is defined. It is furthermore required that, if
$Q(\Gamma)$ is given by
\begin{equation}\label{eq:vn1}
Q(\Gamma)=Q(b;\la) d\la
\end{equation}
according to the representation of $\Gamma$ as a covering of the
$\la$-plane, the closed two-form
\begin{equation} \label{eq:vn2}
\omega_Q=\sum_{i=1}^k d Q(b;\la_i)\wedge d\la_i
\end{equation}
give rise to a Poisson bracket, conveniently called {\em
algebro-geometric\/} Poisson bracket, with $\la_i$ and
$\mu_i=Q(b;\la_i)$ playing the role of Darboux coordinates on the
symplectic leaves of this bracket.
\end{description}

In such a case, it was proven in \cite{VN85} that functions that
depend only on the curve $\Gamma$ -- i.e., on the points of $\CB$
-- are in involution with respect to the Poisson bracket defined
by \rref{eq:vn2},  and these geometric data explicitly define
action-angle variables for these Hamiltonian flows.
In that fundamental paper it has also been shown that a number of
integrable systems, of classical (i.e., mechanical) type as well as  
obtained by suitable reductions of soliton equations, can be framed 
within such a scheme. In particular, the paper \cite{vp00} shows how the
Volterra lattice fits in it. Finally, it is worth mentioning that
Sklyanin's method \cite{Sk95} of the poles of the Baker-Akhiezer
function, originally introduced in the study of Hamiltonian
systems as a byproduct of ``Quantum Integrability", can be seen as
a particularly efficient scheme of implementing the
Veselov--Novikov axiomatic picture.

More recently, a
 \bih\ approach to Separation of Variables (SoV), has
been exposed in the literature (see,
e.g.,~\cite{Bl98,FMT98,FMPZ00,MT97,FP03,BFP04}). Such a scheme
can be seen as a kind of bridge between the classical and the
modern points of view, putting an emphasis on the geometrical
structures of the Hamiltonian theory. In this framework it is
possible to formulate intrinsic conditions on the integrable
system $(H_1,\ldots,H_n)$ to {\em a priori} ensure separability in
a set of canonical coordinates. It  requires  the existence, on
the phase space $(M,\omega)$, of a {\em second\/} Hamiltonian
structure, compatible with the one defined by $\omega$. Namely,
the \bih\ structure on $M$ allows, as it has been shown in a
number of examples,
\begin{enumerate}
\item To encompass the definition of a special set of coordinates, to be
  called {\em Darboux--Nijenhuis (DN)} coordinates, within a well defined
  geometrical object.
\item To formulate intrinsic (i.e., tensorial)
conditions for the separability of a Hamiltonian integrable system,
  in  the DN coordinates associated with the \bih\ structure.
\item To give recipes to characterize, find  and handle sets of
  DN coordinates.
\end{enumerate}

In particular, in \cite{FP03} a detailed discussion of the \bih\
scheme for SoV in the case of Gel'fand-Zakharevich type
\cite{GZ00} systems was presented. It was also pointed out how the
separation relations of such systems were, under genericity
assumptions, of {\em degenerate} type, that is, the functional
form of the separation relations  \rref{eq:1.sjask} is the same
for all pairs of separation coordinates $(\la_i,\mu_i)$, which
essentially means that these coordinates are the coordinates of
different points on the same algebraic curve.

In this paper we want to elaborate further on this issue, and, in
particular, establish a connection between the \bih\ scheme and
the VN setting. This will be done in the first part of the paper,
and, namely, in Proposition \ref{prop:mulax}, which shows that
(under suitable assumptions) DN coordinates associated with
Gel'fand-Zakharevich systems can be seen as algebro-geometric
canonical coordinates in the VN sense. In the second part of the
paper we will apply our scheme in revisiting the
algebro-geometrical integrability \cite{pen98, vp00} of the
well-known Volterra lattice.

More in details, the paper is organized as follows. In Section
\ref{sect:3.1} we will briefly review the \bih\ set--up for SoV.
In Section \ref{sect:3.2}, we will recall some results exposed in
\cite{FP03}, and discuss the relations of the \bih\ approach with
the VN scheme. In Section \ref{sect:4} we will collect a few
results concerning the algebro-geometric scheme of integrating the
Volterra lattice. Finally, in Section \ref{sect:5} we will show
how the \bih\ picture of Section \ref{sect:3.2} can be
successfully applied to the lattice, with no significant
differences between the cases with odd (resp., even) number of
sites.
%% ..... A QUI

\section{Bi-Hamiltonian geometry and Separation of Variables}\label{sect:3.1}

The basic geometrical notion underlying the \bih\ scheme for
separation variables is that of ``semisimple \omnman''. An
\omnman\ is a symplectic manifold $(M,\omega)$ endowed with a
second (possibly degenerate) Poisson tensor $P_1$ which is
compatible with the Poisson tensor $P_0$ associated with the
symplectic form $\omega$. This means that $P_{\lambda}=P_1-\la
P_0$ is a Poisson tensor for all $\lambda\in\mathbb{R}$. In this
case $P_{\lambda}$ is called the Poisson pencil and $(M,P_0,P_1)$
is a \bih\ manifold. It can be shown (see, e.g., \cite{pondi})
that the (1,1) tensor field $N=P_1\circ {P_0}^{-1}$ has the
property
\begin{equation}
  \label{eq:n1}
[N X,N Y]=N\big([N X,Y]+[X,N Y]-N([X,Y])\big),
\end{equation}
for all vector fields $X,Y$ on $M$,
that is, its Nijenhuis torsion vanishes.
The tensor field $N$ is called {\em \Nij\ tensor} or
\emph{recursion operator} of the \omnman. It turns out that
the characteristic polynomial of
$N$ is the square of a polynomial $\Delta(\la)$;
the \omnman\ $M$
is called semisimple provided that
the roots of $\Delta(\la)$ be (generically) simple.

A special class of  coordinates, to be called \Nij\ coordinates,
are provided by the spectral analysis of the adjoint recursion
operator $N^*= P_0^{-1}\circ P_1$. Indeed, one has the following
results (see, e.g., \cite{Ma90,MaMa96,pondi}):
\begin{enumerate}
\item The eigenspace $\Lambda_i$
corresponding to any root $\la_i$ is an integrable
two-dimensional codistribution, that is, one can find $n$ pairs of functions
  $f_i, g_i$ (to be called \Nij\ coordinates) satisfying
  \begin{equation}
    \label{eq:b3}
    N^* df_i=\la_i df_i,\quad N^* dg_i=\la_i dg_i;
  \end{equation}
\item The eigenspaces $\Lambda_i$ and $\Lambda_j$ are
orthogonal with respect to
the Poisson brackets induced both by $P_0$ and $P_1$. This crucial property
can be very simply proven.
Indeed, let $f$ and $g$ be such that their differentials belong respectively
to $\Lambda_i$ and to $\Lambda_j$,
with $i\neq j$. Then one has, on the one hand:
\[
\{f,g\}_1=\langle df, P_1 dg\rangle=\langle N^* df,P_0 dg\rangle=\la_i \{f,g\}_0.
\]
Switching the role of $f$ and $g$ one sees that $\{f,g\}_1=\la_j
\{f,g\}_0$, whence the assertion.
\item Since (as it is easy to
prove) the Poisson bracket $\{f_i,g_i\}_0$ with respect to $P_0$
(and also the one with respect to $P_1$) of functions that
  satisfy~\rref{eq:b3} still satisfies $N^* d \{f_i,g_i\}_0=\la_i \{f_i,g_i\}_0$,
  it is possible to parameterize $\Lambda_i$
with a set of coordinates
  ${x_i,y_i}$, called \dncoo, that are \Nij\ coordinates and are {\em
  canonical} for $\omega$ (whence the addition of
``Darboux'' in their denomination).
\item It is important to notice that the \Nij\ tensor $N$ of an
\omnman\ defines, at each point $m\in M$, a linear operator
$N_m:T_mM\to T_mM$. As such, its eigenvalues (that are point-wise
the roots of $\text{det}(N_m-\la)$) may depend on the point $m$.
In this case we will call these roots, {\em nonconstant} roots of
$\Delta(\la)$. If $\bar\la$ is a nonconstant root of
$\Delta(\la)$, then it satisfies the characteristic
equation\footnote{Actually, the constant eigenvalues trivially
satisfy the same equation.} \rref{eq:b3},
\[
N^* d\bar\la=\bar\la d\bar\la
\]
\end{enumerate}

The following proposition has been proven in \cite{FP03}.
\begin{prop}\label{prop:sepa}
Suppose that $(M,\omega, P_1)$ is a semisimple \omnman\ of dimension $2n$.
Let ${x_1,\dots,x_n,y_1,\dots x_n}$ be
\dncoo\ on $M$ and let $F_1,\ldots,F_n$ be functionally independent
Hamiltonians, that are in involution with respect to the Poisson brackets induced
by $P_0$ and $P_1$. Then the Hamilton-Jacobi equations
associated with any of the Hamiltonians $F_i$ can be solved by additive
separation of variables in the \dncoo\ $(x_i,y_i)_{i=1,\ldots,n}$.
\end{prop}
The foliation given by the functions $F_j$ will be called a {\em
bi-Lagrangian foliation}. Such foliations provide a geometrical
description of separable systems, exactly like Lagrangian
foliations describe integrable systems.

To elaborate further on the geometric structure of \omnman,  it is
convenient to suppose that the eigenvalues of the \Nij\ tensor $N$
be functionally independent. A straightforward observation
\cite{Ma90,FP03}  shows that one can compactly write the
characteristic equation $N^* d\la_i=\la_i d\la_i$ in terms of the
minimal polynomial
\[
\Delta_N(\la)=\prod_{i=1}^{\frac12\text{dim}(M)} (\la-\la_i)
\]
as the polynomial relation
\begin{equation}\label{lem:ladn}
N^* d\Delta_N(\la)=\la d \Delta_N(\la).
\end{equation}
Actually, relations of this kind are very important for our
purposes. Indeed, in \cite{FP03} we proved the following
proposition:
\begin{proposition}\label{prop:nfg}
Let $\Phi(\la)$ be a smooth function defined on the \omnman\  $M$,
depending on an additional parameter $\la$. Suppose that there
exists a one-form $\alpha_\Phi$ such that
\begin{equation}
  \label{eq:2.eeg}
  N^* d\Phi(\la)=\la d\Phi(\la)+\Delta_{N}(\la) \alpha_\Phi\>.
\end{equation}
where $\Delta_N$ is the minimal polynomial of $N$.
Then:\\
a) the $n$ functions $\Phi_i$ obtained evaluating the
``generating'' function $\Phi(\la)$ for $\la=\la_i, i=1,\ldots, n$
are \Nij\ functions,
that is, they satisfy $N^* d\Phi_i=\la_i d\Phi_i$.
\\
b) If $\Phi(\la)$ satisfies \rref{eq:2.eeg} and $Y_l=-P_0 dp_l$
are the vector fields associated via $P_0$ to the coefficients of
the minimal polynomial of the \Nij\ tensor, then all functions
\[
\Phi_l(\la)=\Lie{Y_l} (\Phi(\la)))
\]
satisfy \rref{eq:2.eeg} as well.\\
c) In particular, if $\Phi(\la)$ satisfies, along with
\rref{eq:2.eeg}, the relation
\begin{equation}
\Lie{Y_1}(\Phi(\la))=1 \qquad\text{\rm mod}\,\Delta_{N}(\la),
\end{equation}
then the functions $(\la_i,\mu_i:=\Phi(\la_i))$ provide a set of
\dncoo\ on $M$.
\end{proposition}
\begin{defi} We will call a generating function $\Phi(\la)$ satisfying
  equation~\rref{eq:2.eeg} a {\em \Nij\ functions generator}.
\end{defi}

\section{Gel'fand-Zakharevich systems}\label{sect:3.2}

In many of the models considered in the so-called ``Modern Theory
of Integrable Systems", that is, finite-dimensional Hamiltonian
systems obtained as reductions of integrable PDEs, and/or
classical analogs of quantum spin systems, two instances occur:
\begin{itemize}
\item The phase space $M$ of the system is not the cotangent bundle to a
smooth manifold $Q$.
\item $M$ is endowed with a pair of compatible Poisson brackets,
but none of them is nondegenerate.
\end{itemize}
This is, for instance, the case of the Volterra lattice. This
geometrical instance has been formalized in a series of papers by
Gel'fand and Zakharevich. In particular, under some technical
assumptions, one of which is that the corank of $P_0$ equals that
of $P_1$, it is possible to construct $N=\text{corank}(P_0)$
Lenard--Magri sequences that start with a Casimir function of
$P_0$ and end with a Casimir function of $P_1$. The Hamiltonians
of these sequences can be conveniently collected in polynomials
$\Ha{a}(\la)$, $a=1,\ldots, N$, in the variable $\la$, satisfying
\begin{equation}\label{eq:cpp}
(P_1-\la P_0) d\Ha{a}(\la)=0,
\end{equation}
called {\em polynomial Casimirs} of the {\em pencil}. It is
nowadays customary to denote such a geometrical instance of \varb
s $(M,P_0,P_1)$  with the name of {\em Gel'fand--Zakharevich (GZ)}
manifold.

The collection of the degrees of the polynomial Casimirs is a
numeric invariant of the \varb. Notice in particular that Casimirs
of degree $0$ are nothing but {\em common} Casimir functions of
the two Poisson tensors. We will, in the sequel, call these common
Casimirs {\em trivial Casimirs}, and refer to the others (namely,
those originating a non void Lenard--Magri sequence) as {\em
nontrivial ones}.

In the case of \bih\ systems associated with (coefficients of)
polynomial Casimirs, one can try to use the \bih\ scheme for SoV
reducing the systems to a suitable \omnman. More precisely, one
can consider a symplectic leaf $S\subset M$ of $P_0$ and a
suitable deformation of the Poisson pencil, discussed in detail in
\cite{FP02,FP03,DM02,MB03}.

{}First one fixes a maximal set $C_1, \ldots, C_k$ of independent
nontrivial Casimirs of $P_0$, and finds $k$ (independent)
vector fields $Z_1, \ldots, Z_k$ such that
\[
\Lie{Z_a}(C_b)=\delta_{ab},\qquad \Lie{Z_a} K_\al=0,
\]
where $K_\al$ are the common Casimirs\footnote{In the papers
referred to above, the distribution $\CZ$ spanned by $Z_1, \ldots,
Z_k$ was required to be satisfy the stronger condition $T_pM=T_pS\oplus \CZ_p$ for all $p\in S$.
Actually, as it should be clear form \cite{FP03}, common Casimirs
do not enter the reduction procedure. A similar instance with
common Casimirs has been considered in \cite{F04}.}.
We will hereinafter refer to the
distribution $\CZ$ generated by the $k$ vector fields $Z_i$ as the
{\em transversal distribution}.

Then one considers the vector fields $X_a=P_1 dC_a$, for
$a=1,\ldots,k$, and the ``deformed'' tensor
\begin{equation}\label{eq:pdef}
\widetilde{P_1}=P_1-\sum_{a=1}^k X_a\wedge Z_a,
\end{equation}
that restricts to $S$. If the algebra of functions vanishing along
$Z_1,\ldots,Z_k$ is a Poisson algebra for the pencil $P_1-\la
P_0$, it turns out \cite{FP02} that the deformation
$\widetilde{P_1}$ defines on $S$ a Poisson tensor
$\widetilde{P_1}|S$ compatible with the restriction $P_0|S$ of
$P_0$ to its symplectic leaves. So $S$ is endowed with the
structure of a \omnman. By the definition of $\widetilde{P_1}$,
the restrictions of the coefficients $\Ha{a}_l$ of
$\Ha{a}(\lambda)$ to $S$ will be separable Hamiltonians in the
\dncoo\ defined on $S$ (provided $S$ is semisimple).

{\bf Remark}. To summarize, the ideas underlying such a reduction
procedure are the following: the Gel'fand-Zakharevich scheme
provides -- under some technical assumptions -- a way for
defining, via Magri-Lenard sequences, a distinguished integrable
distribution $\CA$ on a \varb\ $(M, P_0,P_1)$. It is called the
{\em axis} of $M$, and is generated by the Hamiltonian vector
fields associated with the coefficients $\Ha{a}_i$ of the Casimir
polynomials, so that the leaves $\CF$ of $\CA$ are defined by the
requirement that these coefficients be constant along $\CA$.

If we fix our attention on one of the elements of the Poisson
pencil, say, $P_0$, and consider its symplectic foliation $\CS$,
we have that $\CF\cap \CS$
defines, on the generic symplectic leaf $S$ of $\CS$, a Liouville
integrable system.

However, since in general the symplectic foliation associated with
the other Poisson tensor $P_1$ does not coincide with $\CS$, $S$
does not come equipped with a natural \bih\ structure. Finding the
distribution $\CZ$ with the properties outlined above amounts to
finding a deformation $\widetilde{P}_1$ of $P_1$ that

a) endows the symplectic manifold $S$ with a compatible second
Poisson tensor, and hence with the structure of a \omnman.

b) Preserves the commutativity of the Hamiltonians $\Ha{a}_i$,
that is, provides $\CF\cap\CS$ with the structure of a \bilf.

Although to find the transversal distribution $\CZ$ is a non
algorithmic procedure, we notice that this is a quite efficient
way of providing the symplectic leaves $S$ of $P_0$ with a
compatible \Nij\ tensor (a problem which, in principle, requires
the solution of a system of nonlinear partial differential
equations).

\begin{defi}\label{def:affine}
We say that a GZ manifold
$(M,P_0,P_1)$, endowed with a transversal distribution $\CZ$
satisfying the above mentioned assumptions, admits an affine
structure if it is possible to choose a complete set of nontrivial
Casimirs of $P_0$, and a corresponding basis of {\em normalized
flat generators} $\{Z_b\}_\fino{b}{1}{k}$ in $\CZ$ such that, for
every Casimir of the Poisson pencil $H^a(\la)$ and every $b$, $c$,
one has the vanishing of the second Lie derivative of the Casimir
polynomials:
\begin{equation}
  \label{eq:3.x}
  \Lie{Z_b}\Lie{Z_c}(H^a(\la))=0.
\end{equation}
\end{defi}
The above definition might seem somewhat {\em ad hoc}. Its
relevance can be summarized in the following points (whose proof
can be, once more, found in \cite{FP03}) that hold in the case of
affine structures.
\begin{itemize}
\item The nonconstant roots $\la_i$ of the determinant of the
matrix whose entries are
\[
\CG_{ab}=\Lie{Z_b}(H^a(\la))
\]
satisfy
\[
N^* d\la_i=\la_i d\la_i,
\]
namely, they are roots of the minimal polynomial of the \Nij\ tensor induced on
(any of) the symplectic leaves of $P_0$ by the pair $(P_0,
\widetilde{P}_1)$. In particular, if there is only one non-trivial
Casimir polynomial $H(\la)$, and $Z$ is a corresponding normalized
flat generator, the nonconstant roots of the polynomial
$\Lie{Z}(H(\la))$ are ``nonconstant" eigenvalues of the \Nij\
tensor $N$. \item The separation relation satisfied by the
non-trivial Hamiltonian functions and the \dncoo\ are {\em linear}
in the Hamiltonians, that is, are of (generalized) \St\ type.
\end{itemize}

The following Lemma, whose proof is a simple application of some
notions of Poisson geometry, will be frequently used in the
sequel.
It provides a link between the properties of functions on the GZ
manifold $M$, which depend polynomially on the parameter $\la$ of
the Poisson pencil $P_1-\la P_0$ defined on $M$, and the
properties of the evaluation of such functions in $\la=\la_i$
w.r.t. the induced \Nij\ structure on the symplectic leaves. In
plain words, it allows us to work on the GZ manifold $M$, without
having to actually perform the reduction procedure.

We still  suppose that $(M, P_0,P_1)$ is a GZ manifold, with $k$
non-trivial Lenard-Magri sequences. We suppose that $Z_1,\ldots,
Z_k$ are normalized transversal generators for the distribution
$\CZ$ we considered above. We recall that, in this situation, the
symplectic leaves of $P_0$ are \omnman s, with induced \Nij\
tensor $N=P_0^{-1}\widetilde{P_1}.$.
\begin{lemma}\label{lem:rests}
Let $F_\la$ be a function on $M$, invariant along the fields
$Z_i$, that depends holomorphically (say, polynomially) on an
additional parameter $\la$; its restriction $f_\la$ to a
symplectic leaf $S$ of $P_0$ satisfies the ``eigenvector''
equation
\[
N^* df_\la\big\vert_{\la=\la_i}=\lambda_i df_\la\,
\big\vert_{\la=\la_i}
\]
for all eigenvalues $\la_i$  if and only if the following equality
holds, parametrically in $\la$, on the GZ manifold $M$:
\begin{equation}
  \label{eq:n-pb}
\{G,F_\la\}_{P_1}-\sum_{a=1}^k
\Lie{Z_a}(G)\{\Ka{a}_1,F_\la\}_{P_0} =\lambda\{G,F\}_{P_0},
\end{equation}
for any $G\in\> C^\infty(M)$, where the $\Ka{a}_1$ satisfy $P_1
dC_a=P_0 d\Ka{a}_1,\> a=1, \ldots, k$.
Otherwise stated, we have to require that
\[
\widetilde{P_1} dF_\la=\lambda P_0 dF_\la.
\]
where $\widetilde{P_1}$ is defined in \rref{eq:pdef}
\end{lemma}

A direct generalization of the above proposition shows that
spectral curves might be a source for finding \Nij\ functions
generators.

\begin{prop}\label{prop:mulax}
Let us consider a generating function $\Gamma(\la,\mu)$ of
Casimirs of a Poisson pencil $P_\la$, and let us suppose that
$\Gamma(\la,\mu)=0$ defines a smooth algebraic curve. Let $S$ be a
generic symplectic leaf of $P_0$ and let $N$ be the \Nij\ tensor
associated --- according to the scheme outlined above --- with
$P_\la$ and a suitable transversal distribution $\CZ$. Suppose
that $f$ is a $\CZ$-invariant root of the minimal polynomial of
$N$, i.e.,
\begin{equation}
  \label{eq:11}
  \quad N^* df=f df,\quad \text{and } Z_i(f)=0, \fino{i}{1}{k},
\end{equation}
and suppose that $\Gamma(\mu,f)=0$ defines generic point(s) of the
affine curve $\Gamma(\la,\mu)=0$. Then, any solution $g$ of the
equation $\Gamma(g,f)=0$ which is invariant as well under $\CZ$
satisfies $N^* dg=f dg$.
\end{prop}
{\bf Proof:} We first notice the following. Let us consider a
bivariate polynomial $F(\la,\mu)=\sum_{i,j} f_{(i,j)}\la^i\mu^j$,
with coefficients $f_{(i,j)}$ that are functions defined on a
manifold $M$, and two more distinguished functions on $M$, say $f$
and $g$. If we define $\CF:=F(f,g)$, then:
\begin{equation}\label{eq:s12}
d \CF=d F(\la,\mu)\evalc+\ddd{F}{\la}\evalc df+\ddd{F}{\mu}\evalc dg.
\end{equation}
We consider the equation of the ``spectral curve'', $\Gamma(\mu,\la)=0$; so we
get for the zeroes of the function $\CR=\Gamma(f,g)$,
\begin{equation}\label{eq:s13}
0= d \Gamma(\la,\mu)\evalc+\ddd{\Gamma}{\la}\evalc df+\ddd{\Gamma}{\mu}\evalc dg.
\end{equation}
Let us suppose, for simplicity, that
$\Gamma(\la,\mu)=\Gamma_0(\la\mu)+\sum_{i=1}^{k}
\mu^{n_i}H^i(\la)$, where $\Gamma_0(\mu,\la)$ is a constant
polynomial (possibly depending on the common Casimirs), and $n_i$
ares suitable integers. Considering the action of the $k$
transversal vector fields $Z_i$, we get the $k$ equations:
\begin{equation}\label{eq:s14}
\Lie{Z_i}(g)\ddd{\Gamma}{\mu}\evalc+\Lie{Z_i}(f)\ddd{\Gamma}{\la}\evalc+
\sum_{j=1}^{k} \mu^{n_j}\Lie{Z_i}(H^j(\la))\evalc=0.
\end{equation}

Applying $\widetilde{P}_1-f P_0$ to eq.~\rref{eq:s13} we get,
using Lemma~\ref{lem:rests} and taking into account that $(P_1-\la
P_0) d\Gamma(\la,\mu)=0$,
\begin{equation}
  \label{eq:s15}
  \ddd{\Gamma}{\mu}\evalc (N^*-f)dg+ \ddd{\Gamma}{\la}\evalc
  (N^*-f)df-\sum_{i=1}^k \sum_{j=1}^k
  \big(g^{n_i}\Lie{Z_j}H^i(\la)\big)\big\vert_{\la=f}
d H^j_1=0.
\end{equation}
Plugging into equations~(\ref{eq:s14},\ref{eq:s15}) the hypotheses on $f$ we
arrive at the system
\begin{equation}
  \label{eq:s16}
\left\{\begin{split}
  &\Gamma(f,g)=0\\
  &\Lie{Z_i}(g)\ddd{\Gamma}{\mu}\evalc-
\sum_{j=1}^{n} g^{n_j}\Lie{Z_i}(H^j(\la))\big\vert_{\la=f}=0, \quad i=1,\ldots,k.\\
& \ddd{\Gamma}{\mu}\evalc (N^*-f)dg- \sum_{i=1}^n \sum_{j=1}^n
g^{n_i}\Lie{Z_j}H^i(\la)\big\vert_{\la=f} d H^j_1=0.
\end{split}\right.
\end{equation}
The thesis follows noticing that if $g$ is invariant under $\CZ$ then this
system reduces to
\begin{equation}
  \label{eq:s17}
\left\{\begin{split}
  &\Gamma(f,g)=0\\
& \ddd{\Gamma}{\mu}\evalc (N^*-f)dg=0,
\end{split}\right.
\end{equation}
and taking into account that, for $\Gamma(\la,\mu)$ smooth,
the solution of system
\[
\left\{\begin{split}
  &\Gamma(f,g)=0\\
& \ddd{\Gamma}{\mu}\evalc =0
\end{split}\right.
\]
are the (fixed) ramification points of $\Gamma(\la,\mu)=0$.
\endpf

This proposition provides us with the desired link between the
\bih\ approach and the VN axiomatic picture of AG brackets.
Indeed, it can be restated as follows: suppose we can find (by
means of a Lax representation, or by other means) a generating
function for the Casimirs polynomials of an affine GZ pencil in
the form of a bivariate polynomial $\Gamma(\la,\mu)$. Then,
suppose that the coordinates of the points on the curve
$\Gamma(\la,\mu)=0$ (whose $\la$-projections give the roots of the
minimal polynomial of the \Nij\ tensor $N$ induced on symplectic
leaves of $P_0$) satisfy the invariance condition specified in the
above proposition. Then they are \Nij\ coordinates, and so their
Poisson brackets are given by
\[
\{\la_i,\mu_j\}_0=\delta_{ij}
\varphi_i(\la_i,\mu_i),
\quad \{\la_i,\mu_j\}_1=\delta_{ij}
\la_i\varphi_i(\la_i,\mu_i),
\]
Under the further assumption of irreducibility of the minimal polynomial of $N$,
one sees that the unknown functions $\varphi_i
%(\la,\mu)
$ cannot explicitly depend on the index $i$. Thus the formal integral
\begin{equation}
Q(\gamma,\la)d\la=\left(\int^\mu \frac{d\nu}{\varphi(\la,\nu)}\right) d\la
\end{equation}
will give the VN meromorphic differential defining Algebro-geometrical Poisson brackets corresponding to $P_0$
(as well as $Q'=\dsl{\frac{Q}{\la}}$ gives those corresponding to $P_1$).

\section{The Volterra Lattice}\label{sect:4}
The Volterra lattice equations are the following set
\begin{equation} \label{eq:v1}
\dot{c}_i=c_i(c_{i+1}-c_{i-1}),
\end{equation}
which we consider to be defined on a periodic lattice $c_i>0$,
$c_{i+n}\equiv c_i$. They are generalization of the famous
Volterra equations describing time evolution of competing species.

The phase space of the VL can be seen as the restriction to the
submanifold of vanishing momenta of the periodic $n$-site Toda
Lattice. It is well-known \cite{ftbook,pen98,vp00,surbook,Da94}
that eq.~\rref{eq:v1} are isospectral deformation equations for the
periodic difference second order operators of the form
\begin{equation}
(\CL\psi)_k=a_{k+1}\psi_{k+1}+a_k\psi_{k-1},\quad a_{n+i}=a_i, \>
\psi_{n+i}=\la\psi_i. \label{eq:v2}
\end{equation}
where $a_k=\sqrt{c_k}$. In complete analogy with the Toda case, it
admits a {\em dual\/} Lax representation\footnote{This duality
involves also, as in the case of the Toda Lattice, an exchange of
the roles between the spectral parameter $\la$ and the eigenvalue
$\mu$.} in terms of a $2\times 2$ matrix $\CL'$ being given by the
(ordered)
product of site matrices
\[
\CL'=\ell_n\ell_{n-1}\cdots\ell_1, \quad
\ell_i(\mu)=\mat2{\mu}{c_i}{-1}{0}.
\]

The Volterra Lattice equations admit a bi-Hamiltonian formulation.
Indeed, if one considers the quadratic Poisson tensor
\begin{equation}\label{eq:v3p}
P^{ij}=c_ic_j(\delta_{i+1,j}-\delta_{j+1,i})
\end{equation}
and the cubic one
\begin{equation}\label{eq:v3q}
Q^{ij}=c_ic_j(c_i+c_j)(\delta_{i+1,j}-\delta_{j+1,i})
+c_ic_{i+1}c_{i+2}\delta_{i+2,j}-c_{i}c_{i-1}c_{i-2}\delta_{i-2,j}
\end{equation}
one notices that \rref{eq:v1} can be written as
\begin{equation}\label{eq:v4}
\dot{c}_i=\sum_j P^{ij}\ddd{h}{c_j}=\sum_j Q^{ij}\ddd{k}{c_j},
\end{equation}
where
\begin{equation}\label{eq:v5}
h=\frac12\log \prod_{i=1}^{n}c_i,\quad k=\sum_{i=1}^{n} c_i.
\end{equation}
We can collect those remarkable results by Pensko\"\i\ and
Veselov-Pensko\"\i\ \cite{pen98,vp00}, that will be used in the sequel, as follows.

Let us set $\pi=(\prod_{i=1}^{n}c_i)^{1/2}$,
and consider the (normalized) characteristic equation
\begin{equation}
\Gamma(\la,\mu)\equiv \frac{1}\pi\text{Det}(\mu-\CL(\la))=0.
\end{equation}
Then it holds:
\begin{enumerate}
\item $\Gamma(\la,\mu)=\la+\frac1\la-H(\mu)$, where the polynomial
$H(\mu)$ is expressed as
\begin{equation}\label{eq:v6}
H(\mu)=\left\{\begin{array}{l} \sum_{i=0}^k
(-1)^i\mu^{2k+1-2i}\dsl{\frac{J_i}{\pi}}=\\ \\ \quad
\mu\cdot(\sum_{i=0}^k (-1)^i\mu^{2(k-i)}\CH_i),\quad\text{if }
n=2k+1
\\ \\
\sum_{i=0}^{k+1} (-1)^i\mu^{2k+2-2i}\dsl{\frac{J_i}{\pi}}=\\
\\ \quad \mu^2\cdot(\sum_{i=0}^k
(-1)^i\mu^{2(k-i)}\CH_i)+\CH_{k+1},\quad\text{if } n=2k+2
\end{array}
\right.
\end{equation}
The functions $J_i$ can be usefully characterized in the following
way: $J_i$ is the sum of all possible monomials
$c_{l_1}c_{l_2}\cdots c_{l_i}$ of lenght $i$, where the indices
$l_{p}$ are all different and not congruent to $1$ modulo $n$.
In~\cite{vp00} subsets $\{l_1,\ldots,l_i\} \subset\{1,\ldots n\}$
satisfying this property are called {\em totally disconnected} and
we will adopt this definition in the next subsection. Notice that,
e.g., $J_0=1$, $J_1=\sum_i c_i$. Furthermore, notice that for
$n=2k+2$ the last Hamiltonian $\CH_{k+1}$ is given by
\begin{equation}\label{eq:lasth}
\CH_{k+1}=A+\frac{1}{A},\>\text{with } A=\sqrt{\frac{c_2
\,c_4\cdots c_{2k+2}}{c_1\,c_4\cdots c_{2k+1}}}.
\end{equation}
\item \label{pippo}(Theorem 3 of \cite{pen98}). The functions
$\CH_i$ satisfy the Lenard-Magri recursion relations
 \begin{equation}
 P d\CH_0=0,\quad P d\CH_i=Q d\CH_{i-1}, i=1,\ldots , k,\quad Q d\CH_k=0.
 \end{equation}
{}For $n=2k+2$ the last Hamiltonian $\CH_{k+1}$ is in the kernel of
both $P$ and $Q$. \item Let $\{\zeta_1,\ldots,\zeta_{k}\}$ be a
suitably chosen subset of $k$ poles of a suitably normalized
Baker-Akhiezer function $\Psi$ associated with the Lax operator
$\CL$ of \rref{eq:v2},
and let $\la_i$ be corresponding coordinates on the spectral curve
$\Gamma$. Then the coordinates
$\{\zeta_i,\rho_i=\frac{2\log|\la_i|}{\zeta_i}\}$ parametrize the
symplectic leaves of $P$ and satisfy
\begin{equation}
\{\zeta_i,\rho_j\}_P=\delta_{ij},\quad
\{\zeta_i,\rho_j\}_Q=\zeta_i^2\delta_{ij}.
\end{equation}

\end{enumerate}
We remark that one can compactly restate the results of item 2)
in the following form:
\begin{prop}\label{p:t3}
Let us define the Poisson pencil
\[
P_{\mu^2} =Q-\mu^2 P,
\]
where $P$ and $Q$ are the quadratic and cubic Poisson tensors
(\ref{eq:v3p},\ref{eq:v3q}). Then the polynomial $H(\mu)$ defined
by \rref{eq:v6} is a Casimir polynomial of the pencil $P_{\mu^2}$.
Moreover, for $n=2k+2$ the function $\CH_{k+1}$ is a common
Casimir of the two basic elements $P$ and $Q$ of the pencil. Since
functional independence of the Hamiltonians $\CH_i$ is
self-evident, we see that, in the GZ terminology, the phase space
$M$ of the $n$-site Volterra lattice, equipped with the Poisson
structures $P$ and $Q$, is a GZ \varb. If $n=2k+1$ or $n=2k+2$, we
have a single non-trivial Lenard-Magri sequence comprising $k$
vector fields. If $n=2k+2$ is even, the  GZ manifold has a trivial
Casimir $\CH_{k+1}$.
\end{prop}

\section{\SoV\  for the Volterra Lattice in the \bih\ setting}\label{sect:5}
In this section we will make the final contact between the known
results we collected above, and explicitly show how the
picture of \cite{vp00} can be naturally framed within the \bih\ theory
of \SoV\ for GZ systems
described in Section \ref{sect:3.1}

Our starting point are the GZ formulation of Pensko\"i's results, collected in Section 3,
as well as the normalized spectral curve equation~\rref{eq:v6}.
Obviously enough, we will set, as natural parameter of the Poisson pencil,
the quantity $\nu=\mu^2$.

We remark that both in the even and odd number of sites
(or species)
as there is only one non-trivial Lenard--Magri chain,
we have to look for a single transversal vector field to deform the Poisson pencil
$Q-\nu P$.

Let us consider
\begin{equation}\label{eq:zdef}
Z_0=c_1\ddd{}{c_1}+c_n\ddd{}{c_n},
\end{equation}
and define
\begin{equation}
\label{eq:x1}
X=Q d\CH_0.
\end{equation}
The vector field $Z$ is the required ingredient for
applying the \bih\ setting to the Volterra lattice. This follows from the properties
we list and prove below in a series of steps.
\begin{description}
\item{a)} $Z_0$ is a symmetry for the quadratic Poisson tensor $P$,
\[
\Lie{Z_0}{P}=0.
\]
This follows noticing that the Jacobian of $Z_0$ with respect to
the coordinates $(c_1,\ldots,c_0)$ is, in terms of the standard
generators $E_{ij}$ of $n\times n$ matrices, given by
$E_{11}+E_{nn}$, and by the explicit form of $P$. \item{b)} Still
taking this property into account, one can easily verify that the
action of $Z_0$ on $Q$ is given by
\begin{equation}
\Lie{Z_0}Q=Z_0\wedge W_0,
\end{equation}
where
\begin{equation}
W_0=c_1c_n\left(\ddd{}{c_1}-\ddd{}{c_n}\right)+c_1c_2\ddd{}{c_2}-c_{n-1}c_{n}\ddd{}{c_{n-1}}.
\end{equation}
\item{c)}
\begin{equation}\label{eq:Z0ho}
\Lie{Z_0}(\CH_0)=-\CH_0.
\end{equation}
\item{d)} If $n=2k+2$, then
\begin{equation}
\label{eq:b2}
\Lie{Z_0}\CH_{k+1}=0,
\end{equation}
where $\CH_{k+1}$ is given by \rref{eq:lasth}.
These last two properties can be easily verified by straightforward computations.
\end{description}
Hence we can state
\begin{prop}
Let $Z:=-\dsl{\frac{1}{\CH_0}} Z_0$ be the normalized symmetry of
$P$. The two bivectors
\begin{equation}\label{eq:pqdef}
P,\quad \widetilde{Q}=Q-Z\wedge X
\end{equation}
form a Poisson pencil that restricts to the (generic) symplectic leaf
$S$ of $P$.
The functions $\CH_i$ are in involution also with respect to the deformed Poisson bracket $\{\cdot,\cdot\}'$ associated with the bivector $\widetilde{Q}$; hence
their restrictions $\widehat{\CH}_i$ define a bi-Lagrangian foliation of $S$.
\end{prop}
We are now left with the characterization of the \Nij\ coordinates associated with the restriction of the
pencil $\widetilde{Q}-\nu P$ to the symplectic leaf $S$.

Let us consider the vector field $Z$ and the Hamiltonians $\CH_i$,
with $\fino{i}{1}{k}$ if $n=2k+1$, and $\fino{i}{1}{k+1}$ if
$n=2k+2$.
\begin{lemma}\label{lem:ZH}
\begin{equation}
\Lie{Z}(\CH_i)=\Lie{Z}\left(\dsl{\frac{J_i}{\pi}}\right)=J_i\evzer\quad,
\end{equation}
that is, the Lie derivative of $\CH_i$ with respect to $Z$ is
nothing but the numerator of $\CH_i$, evaluated at $c_1=c_n=0$.
\end{lemma}
{\bf Proof}. The proof is a simple chain of computations. We report it here since this Lemma
is crucial for the conclusion of the paper.

We remark that, by the definition of the functions $J_i$ and that of $Z_0$, we have
\begin{equation}
\Lie{Z_0}(J_i)=J_i\Big\vert_{{c_2=\cdots=c_{n-1}=0}}\quad ,
\end{equation}
along with (eq~\rref{eq:Z0ho}) $\Lie{Z_0}(\CH_0)=-\CH_0$. Thus
\begin{equation}\label{eq:Zji}
\Lie{Z}(\CH_j)=-\big(\frac1{\CH_0}\Lie{ Z_0}(\CH_0 J_i)\big)=
-(J_i-J_i\Big\vert_{{c_2=\cdots=c_{n-1}=0}})=J_i\evzer.
\end{equation}
\endpf
{}From the Lemma above, it immediately follows
\begin{prop}\label{pr:minpol}
The polynomial $\Lie{Z}(H(\mu))$ factors as
\begin{equation}
\Lie{Z}(H(\la))=\mu^a\Delta(\nu),
\end{equation}
where $a=1$ if $n$ is odd, $a=2$ if $n$ is even, and $\Delta$ is a
monic degree $k$ polynomial in $\nu=\mu^2$ which, thanks to
\rref{eq:Zji}, is invariant along $Z$.
\end{prop}
{}From the results recalled after Definition \ref{def:affine}, we thus recover the eigenvalues of the
\Nij\ tensor on $S$ as the roots $\nu_i$ of $\Delta(\nu)$.
%%MP Cambiato qui sopra
Since they clearly are functionally independent, we can choose,
as first half of \dncoo, their square roots:
\begin{equation}\label{eq:lacoo}
\zeta_i=\sqrt{\nu_i}.
\end{equation}
Also, if we consider the normalized spectral curve relation, we see that the solutions $\la_i$ of the equation
\begin{equation}
\Gamma(\la,\zeta_i)=0
\end{equation}
are invariant under $Z$ as well, so that choosing one of the
two solutions of this equations will provide a natural candidate for the remaining half of \Nij\ coordinates.

To show that actually this is the case, and simultaneously define a set of \dncoo\  (that is, normalized \Nij\ coordinates), we can rely, once more, on the results of \cite{vp00,pen98}, thanks to the following
\begin{prop}\label{prop:dnpsi}
Let us consider the Lax operator $\CL$ of \rref{eq:v2}  and let
$\psi$ be a Baker Akhiezer vector, for $\CL$, normalized with
$\psi_1=1$. Then on the $k$ points $P_i=(\la_i,\zeta_i)$ chosen
according to the above recipe, $\psi$ has a pole.
\end{prop}
{\bf Proof}. The normalized BA function $\psi$ has poles
in the zeros of the $(1,1)$ element of the classical adjoint matrix
\begin{equation}\label{eq:lad}
(\mu-\CL)^\vee.
\end{equation}
This is the determinant of the $(n-1)\times (n-1)$ matrix
\begin{equation}\label{eq:lad0}
M=\mu{\mathbf
1}-\sum_{a=1}^{n-2}\sqrt{c_{a+1}}(\Ee_{a,a+1}+\Ee_{a+1,a}),
\end{equation}
whose determinant equals the $\Delta(\mu^2)$ if $n$ is odd, and
$\mu\Delta(\mu^2)$ if $n$ is even.
\endpf
This shows that the functions $(\zeta_i,\la_i)_{i=1,\dots,k}$
selected according the \bih\ scheme herewith presented do indeed coincide with
those found by Veselov and Pensko\"\i\ via the method of poles of the BA function.

As a closing remark, we notice that a set of canonical coordinates
for the Volterra lattice can be obtained via ``purely \bih\
methods" as follows. As it has been remarked in Proposition
\ref{prop:nfg}, a possible path is to use the Hamiltonian vector
fields associated with the coefficients $p_i$ of the minimal
polynomial of the \Nij\ tensor $\Delta(\la)$ to deform the
polynomial Casimir of the Poisson pencil. In this way we obtain
new  polynomials that satisfy the characteristic equation of a
\Nij\ functions generator \rref{eq:2.eeg}, and hence we can use
them to generate \dncoo.

{}For the sake of concreteness we will stick to the case of an odd
number of sites $n$.

We consider the minimal polynomial of the induced \Nij\ tensor,
given by (according to Proposition \ref{pr:minpol}, with
$n=2k+1$),
\begin{equation}
\Delta(\nu)=\Big(\frac{1}{\mu}\Lie{Z}(H(\mu))\Big)
\Big\vert_{\mu^2=\nu}=\nu^k-p_1\nu^{k-1}-\cdots p_k, \> \text{where } p_j=-\Lie{Z}(\CH_j).
\end{equation}
Thanks to the explicit characterizations of the Hamiltonians $\CH_i$ and
of the vector field $Z$, and taking Lemma \ref{lem:ZH} into account, it is not
difficult to ascertain that
\begin{equation}\label{eq:pn}
p_k=\prod_{i=1}^k c_{2i}.
\end{equation}
Keeping into account the explicit form of the quadratic Poisson tensor \rref{eq:v3p},
the Hamiltonian vector field associated with $\log{p_k}$ is given by the very simple
expression
\begin{equation}\label{eq:Y}
Y=P\, d\log{p_k}=c_1\ddd{}{c_1}-c_n\ddd{}{c_n}
\end{equation}
If we define as ``first half" of the \dncoo\ the {\em logarithms} of the eigenvalues
of the \Nij\ tensor, rather than their square roots, that is, if we consider
\[
\phi_i=\log(\nu_i),
\]
we clearly have that, in terms of the canonical coordinates $\psi_i$ conjugated
to the $\phi_i$ we are seeking, it holds
\begin{equation}
\label{eq:Y2} Y=P\sum_i d\phi_i=\sum_i \ddd{}{\psi_i},
\end{equation}
so that according to the recipe we are using,
we need to find an exact eigenfunction generator $\Psi(\nu)$ satisfying
\[
Y(\Psi(\nu))\equiv 1\> \text{mod } \Delta(\nu).
\]
Thanks to the explicit form of the vector field $Y$ we can easily establish,
arguing as in the proof of Lemma \ref{lem:ZH}, the following equalities:
%%MP ho cambiato notazione solo qui, mi sembrava che prima appesantisse troppo
\begin{equation}\label{eq:ugu}
\begin{split}
& J^\prime_i:=\Lie{Y}(J_i)=J_i(c_1,0,\dots,0,-c_n),\\
& J^{\prime\prime}_i:=\Lie{Y}(J^{\prime}_i)=J_i(c_1,0,\dots,0,c_n),
\\
& \text{and, finally, } J^{\prime\prime\prime}_i=J'_i.
\end{split}
\end{equation}
This shows that the functions
\[
\eta_i=\log(\nu_i), \quad \psi_i=\log\left(Y(\sum_k \nu_i^k\CH_k)+
Y^2(\sum_k \nu_i^k\CH_k)\right),
\> i=1,\ldots, k.
\]
provide a set of canonical DN coordinates for the Volterra lattice with $2k+1$ sites.
An analogous result holds for the VL with an even number of sites, although the vector
field $Y$ has a more complicated expression.

\section*{Acknowledgments}
We wish to thank Franco Magri for sharing with us his insights about the
problem of SoV.
Some of the results herewith presented
have been obtained in a long--standing
joint-work with him.
We are also grateful to Boris Dubrovin %and Giorgio Tondo
for  useful discussions.
\\
This work has been partially supported by the Italian M.I.U.R.
under the research project {\em Geometric methods in the theory of
nonlinear waves and their applications}, by the ESF project {\em
MISGAM}, and by the European Community through the FP6 Marie Curie
RTN {\em ENIGMA} (Contract number MRTN-CT-2004-5652).

\end{document}